\documentstyle[12pt,aasms4]{article}
\lefthead{Jones et al.}
\righthead{Spectral evidence for dust}
\begin{document}

\title{Spectral evidence for dust in late-type M dwarfs}
\author{Hugh R. A. Jones\altaffilmark{1} and Takashi Tsuji}
\affil{Institute of Astronomy, The University of Tokyo, 2-21-1 Osawa, 
Mitaka 181, Tokyo, Japan}
\authoremail{hjones@mtk.ioa.s.u-tokyo.ac.jp, hraj@staru1.livjm.ac.uk}
\authoremail{ttsuji@mtk.ioa.s.u-tokyo.ac.jp}
\altaffiltext{1}{Astrophysics Group, Liverpool John Moores University, 
Byrom St, Liverpool L3 3AF, UK}

\begin{abstract}
We have found a
dramatic change in the spectra of late-type M dwarfs
at wavelengths shorter than 0.75$\mu$m.
Comparisons with
synthetic spectra illustrate that this change is well explained by the
presence of dust.
It thus seems that a substantial part
of the long-standing discrepancies between late-type M dwarf models
and observations arises from dust as recently proposed by
Tsuji, Ohnaka \& Aoki (1996a).
 
Synthetic spectra and evolutionary models for
low-mass objects $need$ to include dust for the reliable modelling of
objects below 3000 K.
\end{abstract}


\section{Introduction}
 
Although the theory of low-mass stars ($<$ 0.7 M$_{\odot}$) has a long pedigree
(e.g. review Allard \& Hauschildt 1997) models have
had relatively little success in matching observed spectral energy
distributions. It has long been realised that much of this problem may lie
with the lack of good molecular data. However additional reasons have also
been advanced. (1) Brett (1995) propose the neglect of
chromospheric hearing and the simple treatment of convection. (2)
Jones et al. (1995) propose
that the discrepancy with regard to fitting water vapour absorption bands
lies in part with the simplistic treatment of water vapour line
broadening.
(3) Tsuji et al. (1996a) propose that the lack of treatment of dust
opacity has been the root of the problems and showed that
dusty models explain the photometry of the brown
dwarf candidate GD165B (Tsuji et al. 1996b).
 
Here we present comparisons between M dwarf spectra and the predictions
for dusty and non-dusty models. We focus on the region
shortward of 1 $\mu$m where (1) the effects of dust formation can clearly be
distinguished from the problems with reliable modelling of water vapour
absorption bands and (2) the effects of dust formation will be largest, e.g.
fig. 2, Tsuji et al. (1996a).

\section{Observational and synthetic spectra}

Spectra for this paper were taken from a variety of sources
allowing us to check that the effects presented did not arise from
observational errors. 
They come from the
observations of Kirkpatrick, Henry \& McCarthy (1991),
Martin, Rebolo \& Zapatero Osorio (1996) and spectra taken using the FORS
and the RGO spectrograph on the AAT. 
We did find some differences between spectra for objects
common to the different observations though
these were generally smaller than one spectral type. Apart from the object 
BRI0021-0214 (M9.5V) from Martin et al., all the spectra presented are 
Kirkpatrick et al. spectra and thus should be free of significant 
differences due to data reduction procedures.
 
In this paper we discuss the observational spectra in terms of `spectral
type', these exist for large numbers of M dwarfs, are well
quantified and broadly agreed upon.
We might have compared spectra in terms of temperature though, 
the small number
of M dwarfs which have derived temperatures and the large discrepancies
(as much as 600 K) between the relative scales found by different authors
(see discussion Tsuji et al. 1996a) lead us to prefer spectral types.
Traditionally spectral types for
M dwarfs were based on relating TiO and VO band strengths to those
used to define types for M giants, though in recent years there
has been a move to try to base spectral types on the underlying energy 
distribution, rather than the strength of any particular feature.
The spectral types for the objects presented in this paper are
from a least squared minimisation technique (Kirkpatrick et al. 1991)
and from indices based on spectral regions of 'psuedocontinuum' 
(Martin et al. 1996).
Based on comparisons with luminosities and colours these measurements are
primarily sensitive to temperature. Although
there will be some sensitivity to gravity and composition, the objects
presented here are generally standards for their particular spectral type
and are thus expected to have similar gravities and composition.
We thus believe that it is most appropriate to make our comparisons 
considering spectral type as a surrogate for temperature.
 
The spectra examined in this paper
have been presented a number of times by the above authors and others,
however, in Figs \ref{jtobs1} and \ref{jtobs2}
we draw attention to the short-wavelength region
(0.65--0.76 $\mu$m) of the spectra. It can be seen that although the TiO
band strength can be seen to increase between M2V and M6.5V it dramatically
decreases towards later spectral types. This decrease is not expected from
dust-free synthetic spectra and represents the essence of what we are
highlighting in this paper.

The models used for this paper were taken from a grid of models computed
by Tsuji et al. (1996a) to test the effects of grain formation in
low-mass objects. All the synthetic spectra presented here are for solar
metallicity, log $g$ = 5.0 and a microturbulent velocity of 1 km/s.

\section{Spectral evidence for dust}
In Fig. \ref{jtspdiv} we show a spectral sequence made by dividing 
one standard
spectrum by another around two spectral types cooler. The sequence
extends from M2V/M4V to M8V/M9.5V. It can be seen that there is a
dramatic change in the spectra shortward of 0.75 $\mu$m around M7V
and more subtle changes at longer wavelengths around M8V.
This should be compared with the equivalent plot for the synthetic spectra.
Fig. \ref{jtmd} illustrates that around 2500 K the models
predict changes in spectral characteristics 
around 0.67 and 0.71 $\mu$m which are similar in form to Fig. \ref{jtspdiv} 
when dust is included in the models. The feature
at 0.80 $\mu$m is not well reproduced and is due to a lack of reliable
data for VO whose influence is overestimated by the
models (e.g. fig. 12, Brett 1995).

We have investigated
dust-free models from 1000--3800 K with metallicities ranging from
solar to --2 dex and solar metallicity dusty models from 1000--2700 K.
We have not investigated
the effect of changing gravity on the models, however based on the work of
Jones et al. (1996) we expect that gravity differences will be small.
Although there are changes in the characteristics of the spectra
due to temperature (around 1500 and 2300 K) and metallicity we have not
found any combinations of hotter/cooler pairs of dust-free models
which are close to reproducing
the observations. We have only been able to model the change in 
spectral characteristics
around M7V by the inclusion of dust.

In Fig. \ref{jta} we show direct comparisons between observational 
and model spectra from M6V to M9V. 
It can be seen that observational features are reasonably matched
by dusty models. Given the  preliminary treatment of dust employed and 
the uncertainties in the molecular data the fit is a lot better than 
we might expect. 
The observations might also be explained by a combination of inadequacies in
our models:
unaccounted for absorption bands between the TiO and VO bands, 
the presence of a temperature minimum, a chromosphere and an 
overly simplistic
treatment of the molecular opacities. Although it is important to  
investigate these uncertainties we favour an explanation which is
substantially due to dust because with a classical treatment of dust it is
possible to explain the long-standing problem of fitting the observed
water bands (Tsuji et al. 1996a) and at the same time elegantly explain the
behaviour of the short wavelength spectra.

\section{Discussion}

The implication of this work is that dust needs to be considered in models
of objects below 3000 K. This is necessary to enable
a better fit with observed spectra and photometry, for
the derivation of evolutionary tracks for objects below 3000 K
and to find a reliable value for the hydrogen-burning limit.
In addition we caution against the use of
specific molecular bands for assigning spectral types in late-type M
stars.

Given that a dusty model gives a reasonable explanation of
observed spectral differences in late-type M dwarfs further observations
should be made (or possibly dearchived). Spectra for a range
of dwarfs and giants would enable the macroscopic properties of dust to be
constrained at pressures differing by around 10$^6$ across a
range of temperature of perhaps 1800--3000 K. We suggest a number of
observational programmes to constrain dust.

$\bullet$ The models employed in this paper
include dust grains of Al$_2$O$_3$ (corundum), Fe (iron) and
MgSiO$_3$ (enstatite).
Based on fig. 1 (Tsuji et al. 1996a), Al$_2$O$_3$ dust grains are
responsible for most of the changes seen in the spectra presented here.
By careful observations of atomic Al and Mg lines across the transition 
from dust-free to dusty spectra
it should be possible to place constraints on the dust mass and
thus the efficiency of dust formation as a function of temperature, pressure
and abundance.

$\bullet$ Spectra could be investigated for variability  due to dust
absorption thereby constraining the presence of dust clouds in
late-type objects. Based on the spectra that we have
available for vB10 we do find a significant difference between them.
However to reliably test this
hypothesis we feel that it is necessary to use data taken using the same
observational setup and data reduction procedures.

$\bullet$ Spectra at shorter wavelengths than investigated here
are expected to have a stronger dust signature and thus
can thus yield an empirical high
temperature limit to Al$_2$O$_3$ dust formation.

Such tests are not only vital for understanding
the spectral properties of late-type dwarfs stars sufficiently to determine
masses but also to constrain whether radiation pressure on
dust is sufficient to expel matter directly from the photosphere and
is thus the underlying cause of stellar mass-loss in cool giants (e.g. 
Tsuji 1996).

\acknowledgments
It is a pleasure to thank Davy Kirkpatrick and Eduardo Martin for generously
sharing their data.
HRAJ acknowledges PPARC for travel funds.
The work was carried out whilst HRAJ was an EU/JSPS Research Fellow.

\clearpage

\figcaption{Spectra of Gl411 (M2V), Gl268ab (M4.5V) and LHS523 (M6.5V)
are overlayed with their flux normalised to be one at 0.758 $\mu$m.
The deep absorption bands centered around 0.68 and 0.71 $\mu$m
increasing with spectral type are due to TiO.
\label{jtobs1}}

\figcaption{Spectra of LHS523 (M6.5V), vB10 (M8V) and BRI0021-0214 (M9.5V)
are overlayed with their flux normalised to be one at 0.758 $\mu$m.
\label{jtobs2}}

\figcaption{A sequence of `divided' spectra which have been offset
from one another. Working from top down, the uppermost spectra represents
M2V/M4.5V (Gl411/Gl268ab), then M4.5V/M6V (Gl268ab/Gl406), 
M6V/M8V (Gl406/vB10) and M8V/M9.5V (vB10/BRI0021-0214).
\label{jtspdiv}}

\figcaption{Both spectra are a hotter model divided by
a cooler one in the same fashion as the observations in Fig. 3.
In both cases the hotter model is a 2600 K dust-free model;
in the upper plot the cooler model is a 2400 K dust-free model;
in the lower plot the cooler model is a 2400 K dusty model.
\label{jtmd}}

\figcaption{A comparison between observed spectra (solid), dusty models
(dashed) and dust-free models (dotted): (a) Gl406(M6V) and 2600 K models, 
(b) vB10 (M8V) and 2200 K models, (c) LHS2924 (M9V) and 2000 K models.
\label{jta}}

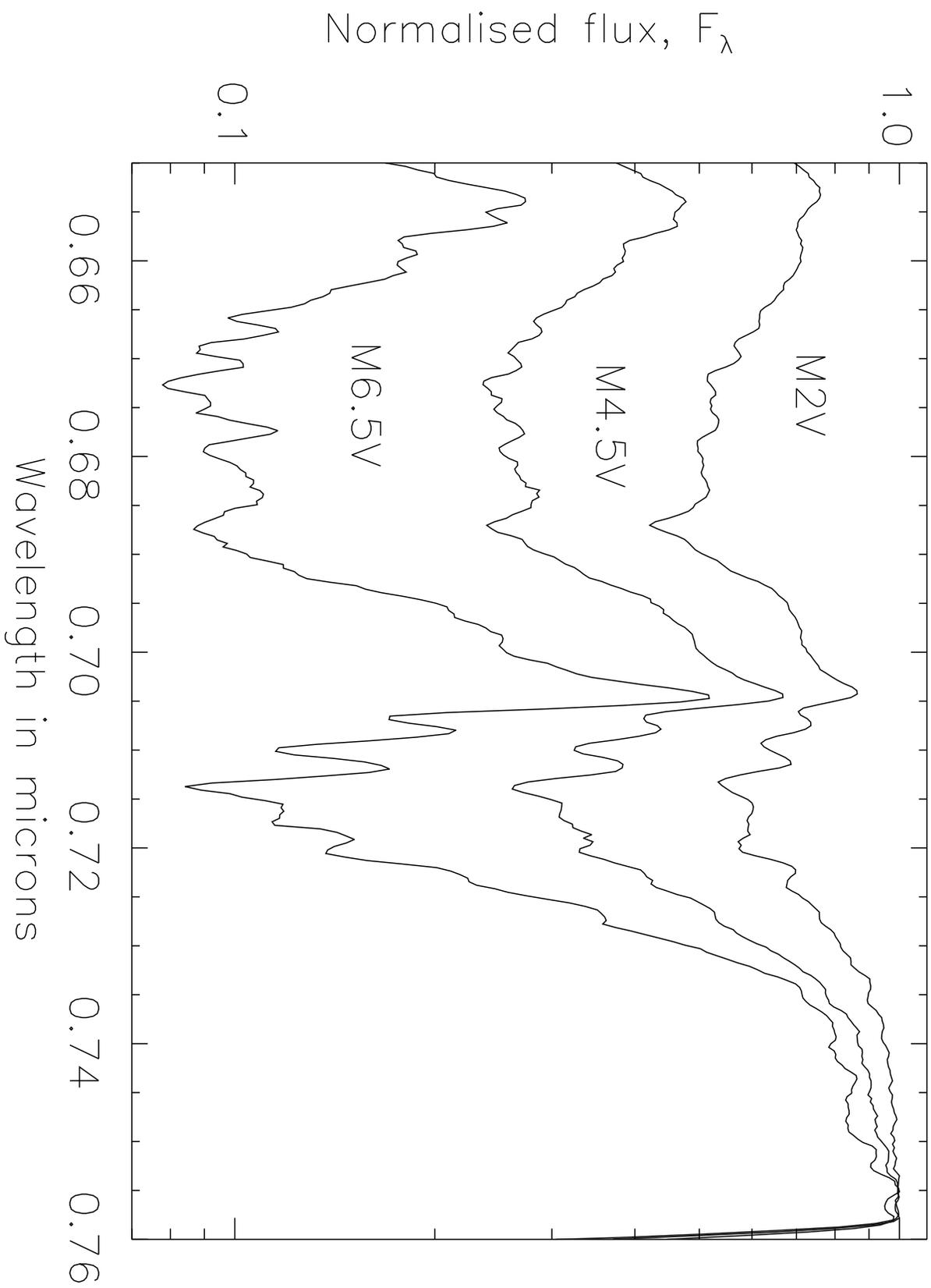
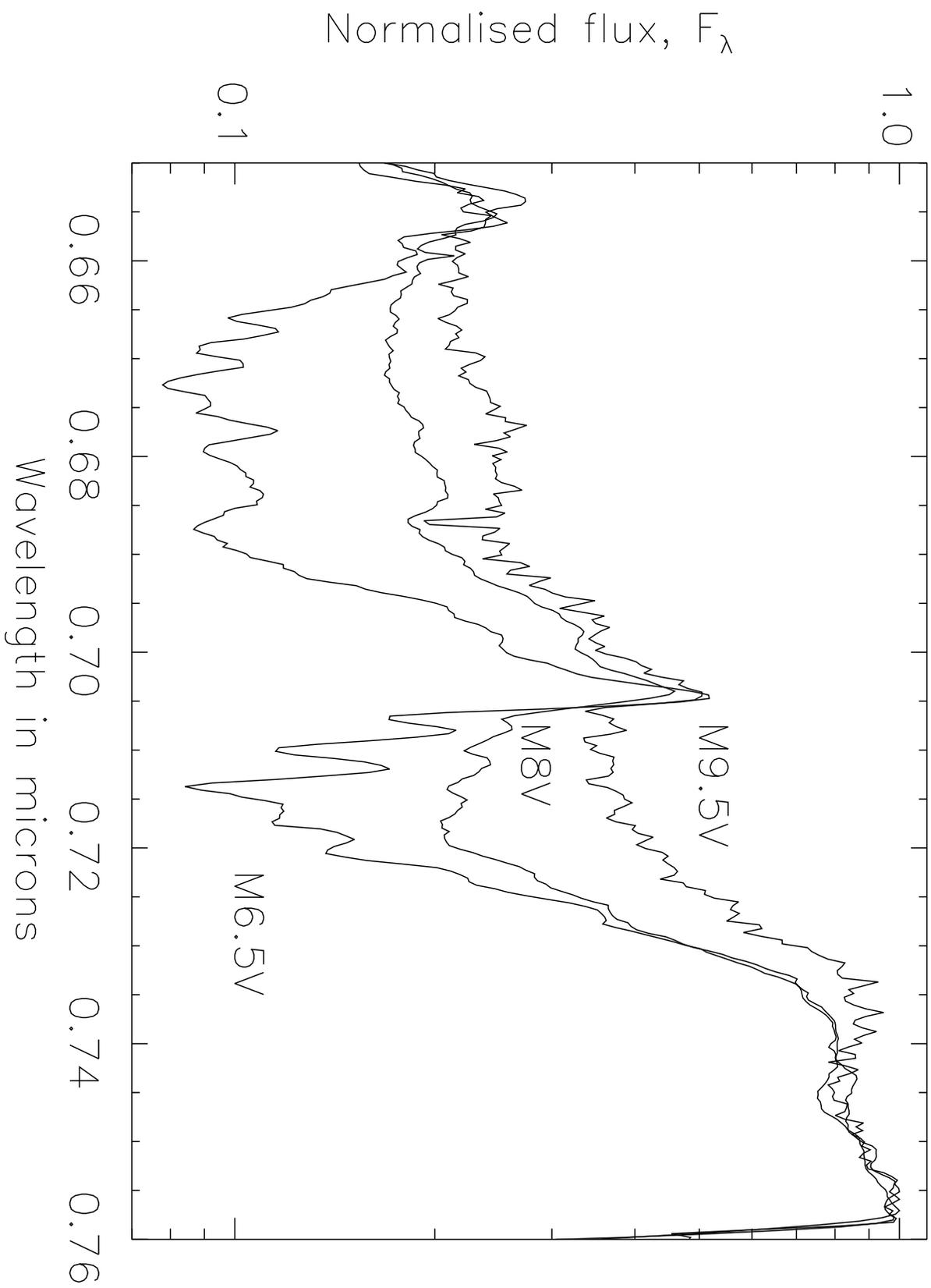
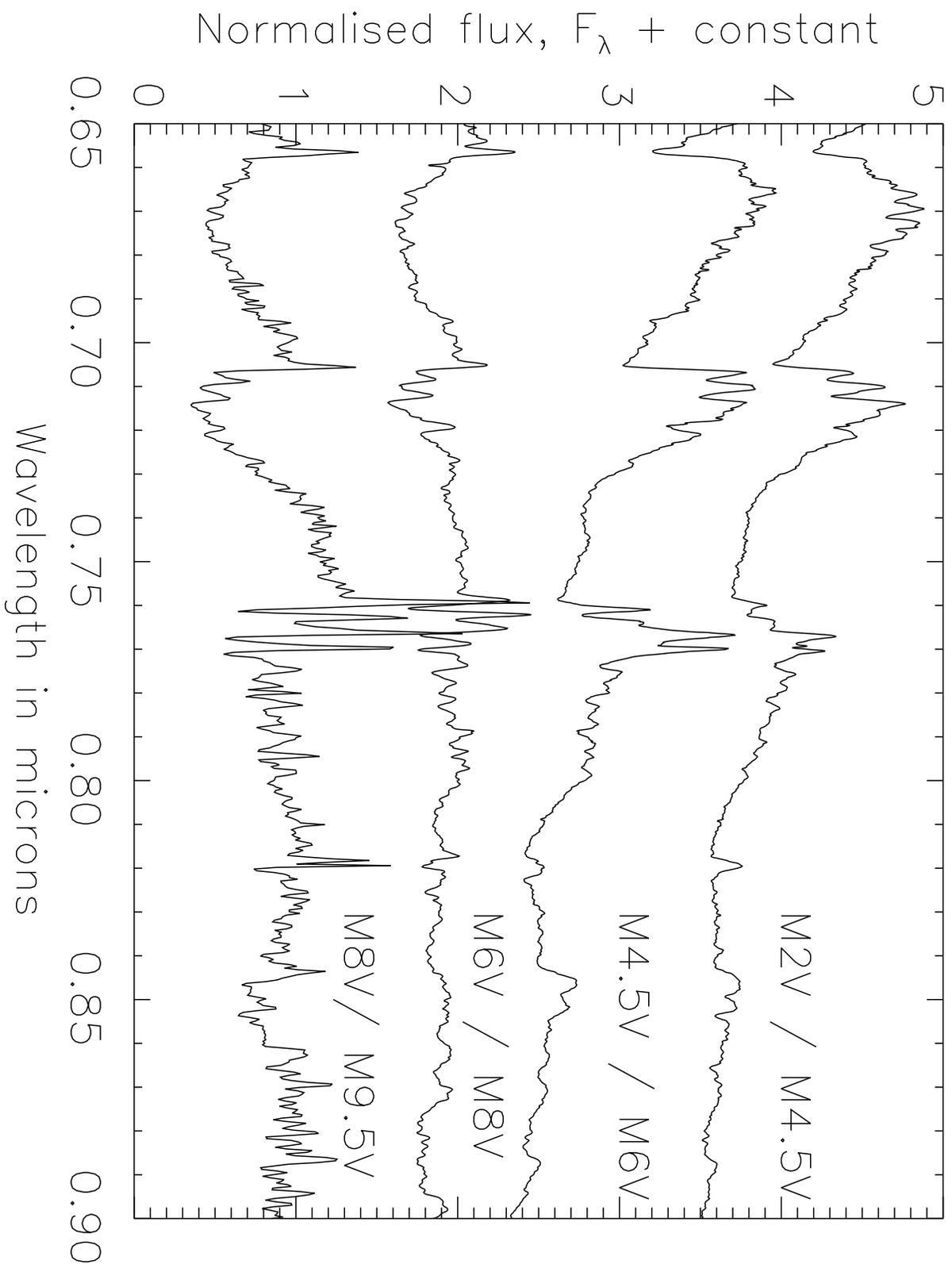
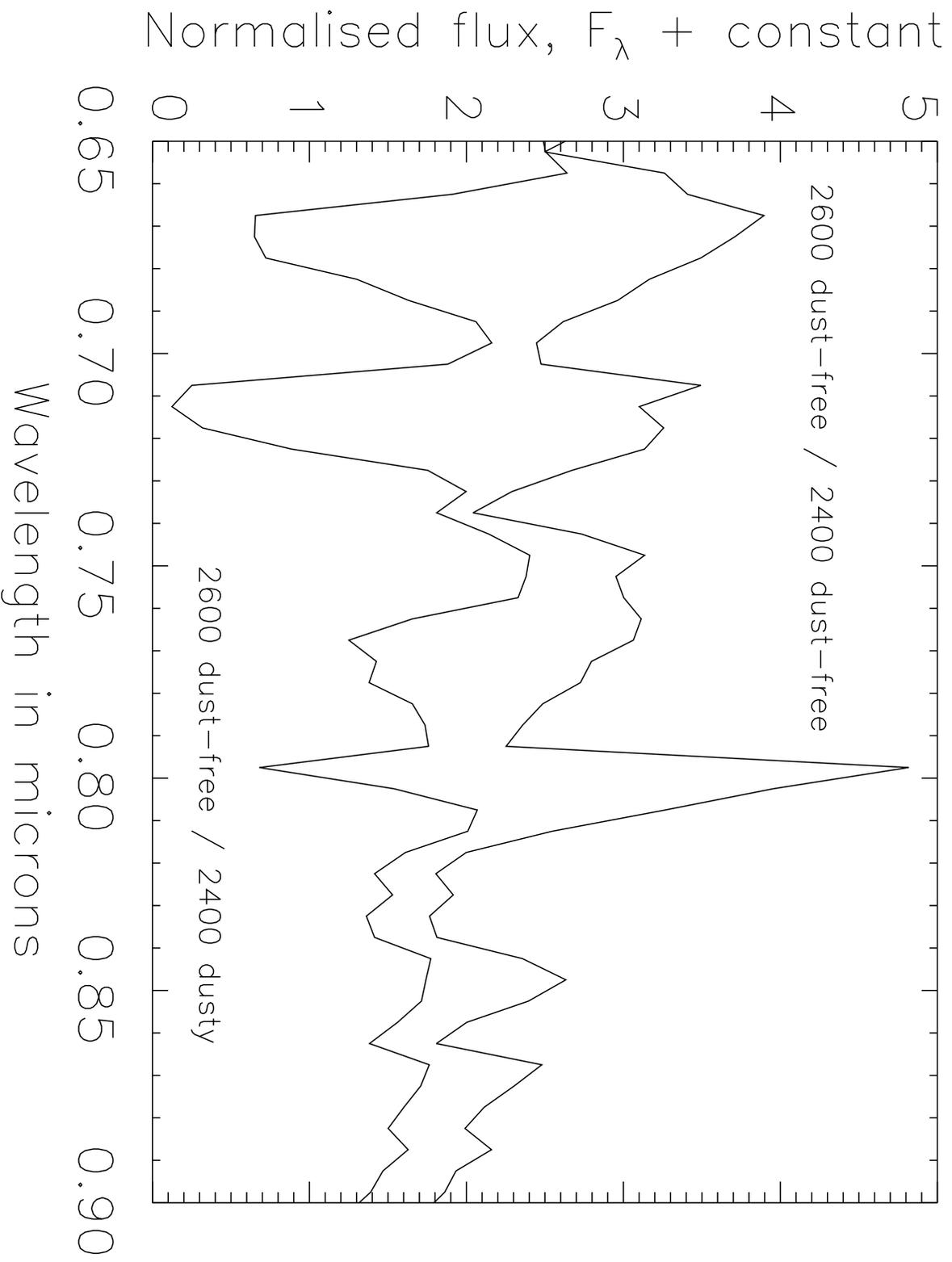
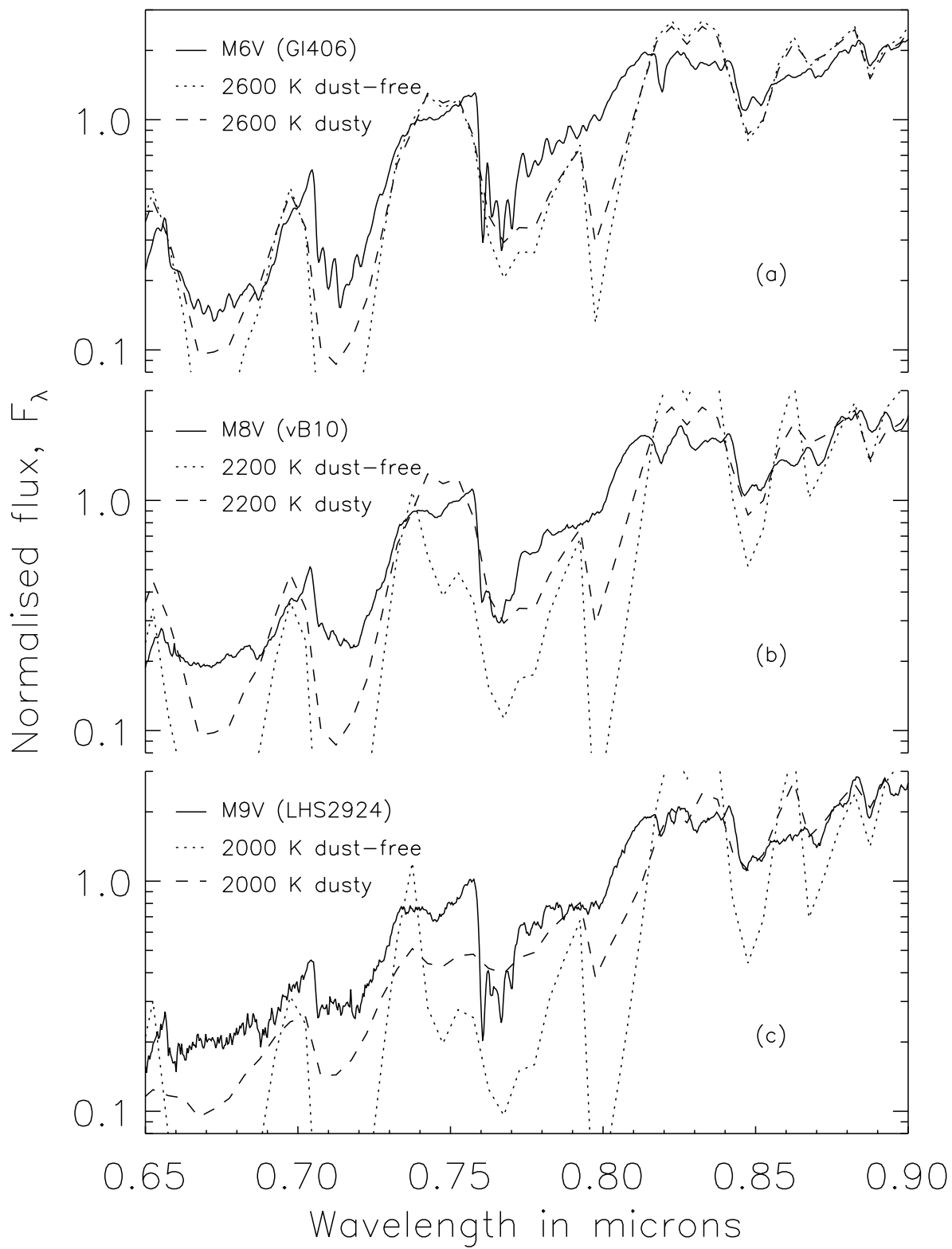
\end{document}